\documentclass[american,aps,pra,reprint,superscriptaddress,onecolumn]{revtex4-1}
\usepackage{amsmath,amscd,amsthm,amssymb}
\usepackage{mathrsfs}
\usepackage{graphicx}
\usepackage{epsf,epstopdf}
\usepackage[all]{xy}
\usepackage[unicode=true,pdfusetitle, bookmarks=true,bookmarksnumbered=false,bookmarksopen=false, breaklinks=false,pdfborder={0 0 0},backref=false,colorlinks=false] {hyperref}
\hypersetup{colorlinks=true,linkcolor=myurlcolor,citecolor=myurlcolor,urlcolor=myurlcolor}
\usepackage{graphics,epstopdf,graphicx, amsthm, amsmath, amssymb, times, braket, color, bm}
\usepackage{cleveref}
\usepackage{caption}
\usepackage{subcaption}
\usepackage{makecell}

\definecolor{myurlcolor}{rgb}{0,0,0.7}

\theoremstyle{plain}

\providecommand{\theoremname}{Theorem}
\newcommand*{\myproofname}{Proof}

\makeatother

\theoremstyle{definition}

\theoremstyle{remark}

\begin{document}

\title{Arbitrary state creation via controlled measurement}

\author{Alexander I. Zenchuk}
\email{zenchuk@itp.ac.ru}
\affiliation{Federal Research Center of Problems of Chemical Physics and Medicinal Chemistry RAS,
Chernogolovka, Moscow reg., 142432, Russia}

\author{Wentao Qi}
\email{qiwt5@mail2.sysu.edu.cn}
\affiliation{Institute of Quantum Computing and Computer Theory, School of
Computer Science and Engineering,
Sun Yat-sen University, Guangzhou 510006, China}


\author{Junde Wu}
\email{wjd@zju.edu.cn}
\affiliation{School of Mathematical Sciences, Zhejiang University, Hangzhou 310027, China}


\begin{abstract}
The initial state creation is a starting point of many quantum algorithms and usually is considered as a separate subroutine not included into the algorithm itself. There are many algorithms  aimed on creation of special class of states.
Our  algorithm allows  creating  an arbitrary $n$-qubit pure quantum superposition state with  precision of $m$-decimals (binary representation) for each probability amplitude.  The algorithm uses one-qubit rotations, Hadamard transformations and C-NOT operations with multi-qubit controls. However, the crucial operation is the final  controlled measurement of the ancilla state that removes the garbage part of the superposition state and allows to avoid the problem of small  success probability in that measurement.
We emphasize that rotation angles are predicted in advance by the required precision  and therefore there is no classical calculation supplementing quantum algorithm. The depth and space of the algorithm growth with $n$  as, respectively,
$O(2^n  n)$
and  $O(n)$. 
This algorithm can be a subroutine generating the required input state in various algorithms, in particular,  in  matrix-manipulation algorithms developed earlier.
\end{abstract}

\maketitle

{\it Introduction.}
The quantum state preparation is a vital process in quantum informatics having  both theoretical and experimental aspects \cite{NCh,WMC}.
In many cases this step is not  included into the main algorithm which assumes  access to the  already prepared quantum state.  However, namely the initial state creation might be the most crucial  in determining the characteristics of the  algorithm, such as depth and space, thus governing  the effectiveness of the algorithm.
The particular initial state preparation is required, for instance, in the HHL-algorithms solving system of linear algebraic equations \cite{HHL} to encode the known rhs of the linear system,  in the algorithm of matrix manipulation to create the matrices to be subjected to addition, multiplication, inversion \cite{ZZRF,ZQKW_2024,ZBQKW_arXive2024},  in quantum machine learning to encode the input data \cite{BWPRWL,A,RML}, in  the least-square linear-regression algorithms  \cite{WBL,SSP,Wang2} working with encoding the  large data sets.

The known algorithms of state creation are usually well suitable  for creating certain classes of states.
Thus, in \cite{MVBS,BVMS}, the uniformly controlled rotations  are  used for transforming the input state to the required form. Although the algorithm is presented as a rather universal one its realization requires special calculation of needed rotation angles via  classical tool. In \cite{APPS},  the   divide-and-conquer algorithm \cite{CLRS} was used to speed up the data loading.
Arbitrary state preparation in the Schmidt decomposition form is considered in \cite{PB} and is suitable for creating the small-scale states.
The preparation of  quantum states
 that
are uniform superpositions over a subset of basis states is considered in \cite{MSRM,SV}.  These algorithms have advantages in characteristics but suffer in variety of creatable states.
The algorithm for preparing the  Qudit Dicke states (equal-weight superposition of all states with fixed number of excited qubits)  is presented in \cite{NR} and also suffer from restrictions on the family of creatable states.
Quantum netwoks  \cite{KM} may be effective for high-fidelity preparing,  in particular, pure symmetric states.
The $O(n)$-depth algorithm encoding the $n$-qubit state using exponential amount of ancillary qubits is discussed in \cite{ZLY}.  Obviously,  its weak point is the required space.
A method for encoding  vectors obtained by
sampling analytical functions into quantum circuits is proposed in \cite{MTDNP}. It has privilege in reaching high fidelity  but needs a particular function whose simpling yields the required vector (state).
The protocol incorporating
periodic quantum resetting for
preparing ground states of frustration-free
parent Hamiltonians is studied in \cite{PMCMR}.
The ground state of the spin-1 Affleck, Kennedy, Lieb and Tasaki (AKLT) model can be prepared
 using fusion  measurements (fusion of small matrix product
states through the Bell measurement of a special ancilla)  \cite{SCWG}.
The algorithms discussed in \cite{NDW,LMDW,DFZ_2020} require utilizing specific unitary transformations with particular rotation parameters determined via classical computation and therefore  require classical supplementary calculations. Large variety of quantum states can be created via adiabatic technique \cite{WMC,AL}.

In our paper, we represent the special algorithm allowing to create an arbitrary $n$-qubit  pure quantum state
\begin{eqnarray}\label{tPsi}
|\tilde \Psi\rangle =  \sum_{j=0}^{2^n-1} \tilde a_j e^{2 \pi i \tilde \varphi_j}  |j\rangle_S,\;\;\sum_{j=0}^{2^n-1}  \tilde a_j^2=1,
\end{eqnarray}
of the system $S$ via one-qubit rotations, Hadamard operators and the set of C-NOTs with multi-qubit control. We use $m$-decimal approximation (in binary form) of both amplitudes $\tilde a_j$ and phases  $\tilde \varphi_j$.  To  encode the approximate amplitudes and phases   we  involve  two $m$-qubit subsystems.
We show that the  depth  of  the algorithm is $O(2^n nm + m^2)\sim O(2^n n)$ and the space is $O(n+m)\sim O(n)$ qubits.
Although the depth is large, the advantage of our algorithm  is that  it doesn't assume any additional calculations of the parameters for the rotations  except the  binary  expansion of the amplitudes and phases of the state to be  encoded. In other words, it doesn't require any additional inclusion of classical computations.
 The principal issue of this algorithm is the controlled measurement \cite{FZQW_arxive2025} that, first of all, removes the garbage and second (that is more important)  it removes the problem  of small success probability in the mentioned ancilla measurement that necessarily appears if we use usual measurement instead of the controlled one.

{\it Arbitrary state preparation.
\label{Section:InSt}
Let us consider  an arbitrary $n$-qubit quantum state (\ref{tPsi}),
which is to be encoded into the state of the quantum system $S$.
In (\ref{tPsi}), all amplitudes $\tilde a_j$  and phases $\tilde\varphi_j$ are real numbers with $0\le \tilde a_{j} \le 1$, $0\le \tilde \varphi_{j} < 1$.

Of course, we can not encode the exact state (\ref{tPsi}) in general because we must  fix the precision used for representations amplitudes $\tilde a_j$ and phases  $\tilde\varphi_j$. Therefore,  we encode  an approximate state prepared as follows. Let us  approximate  $\tilde a_j$ and $\tilde \varphi_j$ keeping $m$ decimals in the binary form, i.e.,
\begin{eqnarray}\label{tildea}
&&
\tilde a_j \approx \sum_{k=1}^{m} \frac{\alpha_{j(m-k)}}{2^k}=\frac{1}{2^{m}}  \sum_{k=0}^{m-1}
\alpha_{jk} 2^k=\frac{a_j}{2^{m}},\;\; a_j =\sum_{k=0}^{m-1}
\alpha_{jk} 2^k  ,\\\nonumber
&&\tilde \varphi_j \approx \varphi_j = \sum_{k=1}^{m} \frac{\beta_{jk}}{2^k},
\end{eqnarray}
where all $\alpha_{jk}$ and   $\beta_{jk}$ equal either 0 or 1.
This allows us to replace the state (\ref{tPsi}) with the approximate one
\begin{eqnarray}\label{Psi}
| \Psi\rangle = G^{-1} \sum_{j=0}^{2^n-1} a_j e^{2 \pi i \varphi_j}  |j\rangle,\;\;G=\sqrt{ \sum_{j=0}^{N-1} a_j^2}.
\end{eqnarray}
Below we discuss the quantum algorithm encoding the approximate state $|\Psi\rangle$ using $\alpha_{jk}$ and   $\beta_{jk}$ as parameters in the controlled operations included into the algorithm.
To create  the state $|\Psi\rangle$ given  in  (\ref{Psi}) we involve the $n$-qubit subsystem $S$ which stores the state $|\Psi\rangle$ and two auxiliary $m$-qubit subsystems $R$ and $\varphi$,
those are responsible for the accuracy of creation of, respectively,  the amplitudes $a_j$ and phases $\varphi_j$  in the  state-creation algorithm.

We  start with the ground state of the subsystems $S$, $R$ and $\varphi$:
$|\Phi_0\rangle=|0\rangle_S |0\rangle_R|0\rangle_\varphi$,
see Fig.\ref{Fig:instgen}.
\begin{figure}[ht]
  \includegraphics[width=1\textwidth]{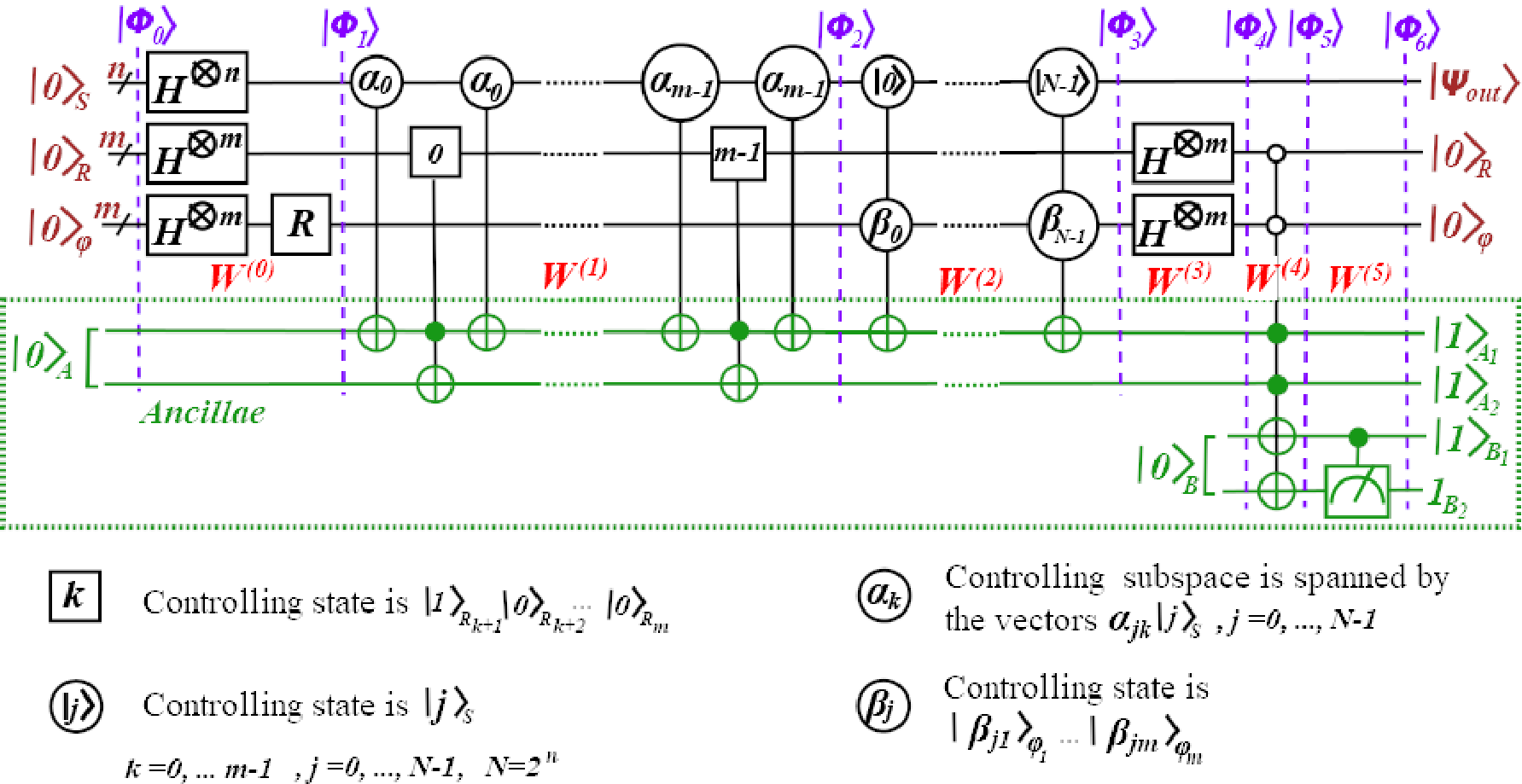}
    \caption{The circuit for creating an arbitrary quantum state, $R=\prod_{k=1}^m R_{\varphi_k}$.}
\label{Fig:instgen}
\end{figure}
First, we apply the Hadamard transformations $H_S=H^{\otimes n}$,  $H_R=H^{\otimes m}$ and $H_\varphi=H^{\otimes m}$, $H=1/\sqrt{2}\left( \begin{array}{cc}1&1\cr
1&-1\end{array}\right)$,   to each qubit of the subsystems $S$, $R$ and $\varphi$ respectively  and, in addition, apply the gate
\begin{eqnarray}
R_{\varphi_k} = \left(
\begin{array}{cc}
1&0\cr
0 &e^{2\pi i /2^k}
\end{array}
\right), \;\;k=1,\dots,m,
\end{eqnarray}
to  the $k$th  qubit of the subsystem $\varphi$.  Thus we form the
operator   \begin{eqnarray}
 W^{(0)}_{SR\varphi}=\prod_{k=1}^m  R_{\varphi_k} H_{\varphi} H_RH_S.
 \end{eqnarray}
Hereafter, the subscript at the operator indicates the subsystem to which this operator is applied.  The subscript at the subsystem indicates its  qubit.
 Applying $W^{(0)}_{SR\varphi}$  to the state
 $|\Phi_0\rangle$ we obtain the state $|\Phi_1\rangle$, see Fig.\ref{Fig:instgen}:
\begin{eqnarray}\label{Phi1}
|\Phi_1\rangle=W^{(0)}_{SR\varphi} |\Phi_0\rangle  =\frac{1}{2^{(n+2m)/2}}
\sum_{j=0}^{2^n-1} |j\rangle_S \sum_{k=0}^{2^m-1} |k\rangle_R |\Psi_{\varphi }\rangle
\end{eqnarray}
where we denote, for brevity,
\begin{eqnarray}
|\Psi_{\varphi }\rangle =
\prod_{k=1}^m \Big(|0\rangle_{\varphi_k} + e^{2\pi i /2^k} |1\rangle_{\varphi_k} \Big).
\end{eqnarray}
After such preparations we propose two separate  subroutines encoding the amplitudes $a_j$ 
and phases $\varphi_j$ 
in the superposition state $|\Psi\rangle$ of the subsystem $S$. 
In addition, each subroutine  involves one additional $m$-qubit subsystem mentioned above, either  $R$ (amplitude encoding) or $\varphi$ (phase encoding).

{\it Amplitude encoding.}
First, we encode  the integer} amplitudes $a_j$ in the state $|\Psi\rangle$ given in  (\ref{Psi}).
We perform this encoding relating $a_j$ different basis states of the subsystem  $R$ from the sum $\sum_{k=0}^{2^m-1} |k\rangle_R$ to the state $|j\rangle_S$.
For this purpose we introduce the 2-qubit ancilla $A$ in the ground state $|0\rangle_A$, the projectors
\begin{eqnarray}\label{P2}
P^{(k)}_{S} = \sum_{j=0}^{2^n-1}  {\alpha_{jk}}  |j\rangle_S\,_S\langle j| ,\;\;k=0,\dots, m-1,
\end{eqnarray}
and the appropriate   controlled operators
\begin{eqnarray}\label{tW1}
\tilde W^{(k)}_{SA_1} = P^{(k)}_{S} \otimes \sigma^{(x)}_{A_1} +(I_S- P^{(k)}_{S} )\otimes I_{A_1}  ,\;\;k=0,\dots, m-1.
\end{eqnarray}
The operator $\tilde W^{(k)}_{SA_1}$ with the fixed $k$ relates  the terms  $\alpha_{jk}|j\rangle_S$ in
$|\Phi_1\rangle$, $j=0,\dots,2^n-1$, for which $\alpha_{jk}\neq 0$,  to the excited state of the ancila qubit $A_1$,
$|1\rangle_{A_1}$.
Now we introduce another set of projectors
\begin{eqnarray}\label{P1}
P^{(k)}_{RA_1} =|1\rangle_{R_{k+1}}|0\rangle_{R_{k+2}} \dots|0\rangle_{R_{m}}  \, {_{R_{k+1}}\langle 1|} {_{R_{k+2}}\langle 0|}  \dots {_{R_{m}}\langle 0|}   \otimes  |1\rangle_{A_1} {_{A_1}\langle 1| },\;\;k=0,\dots, m-1,
\end{eqnarray}
and the control
 operators
 \begin{eqnarray}\label{W1}
V^{(k)}_{RA}=P^{(k)}_{RA_1}\otimes \sigma^{(x)}_{A_2} + (I^{(k)}_{RA_1} - P^{(k)}_{RA_1})\otimes I_{A_2} , \;\;k=0,\dots, m-1,
\end{eqnarray}
where $I^{(k)}_{RA_1}$ means the identity operator applied to the last  $m-k$ qubits of the subsystem $R$ and to
the ancilla  qubit $A_1$. The operator $V^{(k)}_{RA}$ with  fixed $k$  relates $2^{k}$ basis states of $R$
with each state in the sum $\sum_{j=0}^{2^n-1}|j\rangle_S$ in $|\Phi_1\rangle$, for which $\alpha_{jk}\neq 0$,  through the excited state of the ancilla qubit $A_2$, $|1\rangle_{A_2}$. The information about the amplitudes  to be
created (parameters $\alpha_{jk}$ from expansions (\ref{tildea})) is enclosed  into the projectors $P^{(k)}_S$ given in (\ref{P2}) and does not
require additional calculations via classical tool.
Collecting formulae (\ref{P2})-(\ref{W1}) we   construct the operator
\begin{eqnarray}\label{W12}
&&
W^{(1)}_{SRA} = \prod_{k=0}^{m-1}  \tilde W^{(k)}_{SA_1} V^{(k)}_{RA} \tilde W^{(k)}_{SA_1}.
\end{eqnarray}
Here, the third operator $\tilde W^{(k)}_{SA_1} $ in the triad
$\tilde W^{(k)}_{SA_1} V^{(k)}_{RA} \tilde W^{(k)}_{SA_1}$ returns the ancilla qubit $A_1$ to the ground state $|0\rangle_{A_1}$ thus preparing this qubit for using by the next triad of operators  $\tilde W^{(k+1)}_{SA_1} V^{(k+1)}_{RA} \tilde W^{(k+1)}_{SA_1}$.
Applying  the operator $W^{(1)}_{SRA}$ to the state $ |\Phi_1\rangle |0\rangle_{A}$  we obtain the state
$|\Phi_2\rangle$, see Fig.\ref{Fig:instgen}:
\begin{eqnarray}\label{Phi2}
&&
|\Phi_2\rangle = W^{(1)}_{SRA} |\Phi_1\rangle |0\rangle_{A}   =\\\nonumber
&&
\frac{1}{2^{(n+2m)/2}}
\sum_{j=0}^{2^n-1} |j\rangle_S \sum_{k=0}^{m-1} \alpha_{jk}\sum_{l=0}^{2^k-1} |2^k+l\rangle_R  |\Psi_{\varphi }\rangle   |0\rangle_{A_1} |1\rangle_{A_2} +|g_2\rangle.
\end{eqnarray}
In (\ref{Phi2}), the first part collects the terms with the excited state of $A_2$, while all other terms are collected in the garbage $|g_2\rangle$ to be removed  later. All together, each state $|j\rangle_S$ is related  to $\sum_{k=0}^{m-1}\alpha_{jk} 2^{k}$ basis states  of the subsystem $R$.

Thus, the information about the amplitudes of the state $|\Psi\rangle$ is encoded   into the state $|\Phi_2\rangle$ through the parameters $\alpha_{jk}$. Before complete the  amplitude encoding we turn to  the phase encoding algorithm.

{\it Phase encoding.} The phase encoding subroutine differs from the subroutine encoding the amplitudes $a_j$.
To encode the phases $\varphi_j$ into the state $|\Phi_2\rangle$ (and eventually into the state $|\Psi\rangle$ given in (\ref{Psi})) we introduce the   projectors
\begin{eqnarray}\label{P22}
P^{(j)}_{S\varphi} =|j\rangle_S\,_S\langle j|   \prod_{k=1}^m  |\beta_{jk}\rangle_{\varphi_k}\, {_{\varphi_k}\langle \beta_{jk} |} ,\;\;j=0,\dots,  2^n-1,
\end{eqnarray}
and the controlled operators
\begin{eqnarray}\label{W2SA}
\tilde W^{(j)}_{S\varphi A_1} = P^{(j)}_{S\varphi} \otimes \sigma^{(x)}_{ A_1} +  (I_{S\varphi}- P^{(j)}_{S\varphi}) \otimes I_{A_1},\;\;j=0,\dots,  2^n-1.
\end{eqnarray}
The complete information about the phases (parameters $\beta_{jk}$)  is encoded into the projectors $P^{(j)}_{S\varphi}$ given in (\ref{P22}).
Now we construct the operator
\begin{eqnarray}\label{W2SA2}
W^{(2)}_{S\varphi A_1} =\prod_{j=0}^{2^n-1}  \tilde  W^{(j)}_{S\varphi A_1}
\end{eqnarray}
and apply it  to the state $ |\Phi_2\rangle$, thus obtaining the state $|\Phi_3\rangle$, see Fig.\ref{Fig:instgen},
\begin{eqnarray}\label{Phi3}
&&
|\Phi_3\rangle = W^{(2)}_{S\varphi A_1} |\Phi_2\rangle = \\\nonumber
&&
\frac{1}{2^{(n+2m)/2}}
\sum_{j=0}^{N-1} |j\rangle_S \sum_{k=0}^{m-1} \alpha_{jk}\sum_{l=0}^{2^k-1} |2^k+l\rangle_R \prod_{r=1}^m e^{2\pi i \beta_{jr} /2^r} |\beta_{jr}\rangle_{\varphi_r} |1\rangle_{A_1}  |1\rangle_{A_2}    +  |g_3\rangle.
\end{eqnarray}
We see that the operator $W^{(2)}_{S\varphi A_1}$ serves to create the needed phases expressed in terms of $\beta_{jr}$ so that $\varphi_j = \sum_{k=1}^m \frac{\beta_{jk}}{2^k}$.
Next, we identify all states $|l\rangle_R$ and all states $|\beta_{jr}\rangle_{\varphi_k}$ in (\ref{Phi3}).  To this end we
apply the Hadamard transformations $H_R=H^{\otimes m}$ and $H_\varphi=H^{\otimes m}$ to each qubit of
the subsystems $R$ and $\varphi$, i.e., the transformation
\begin{eqnarray}\label{W3}
W^{(3)}_{R\varphi}= H_RH_\varphi,
\end{eqnarray}
and select the terms with the state $|0\rangle_R|0\rangle_\varphi  |1\rangle_{A_1}  |1\rangle_{A_2}$, putting other terms into the garbage
$|g_4\rangle$:
\begin{eqnarray}\label{Phi4}
|\Phi_4\rangle = W^{(3)}_{R\varphi} |\Phi_3\rangle =\frac{1}{2^{(n+4m)/2}}  \sum_{j=0}^{2^n-1}
a_je^{2\pi i\varphi_j} |j\rangle_S |0\rangle_R|0\rangle_\varphi  |1\rangle_{A_1}  |1\rangle_{A_2}   + |g_4\rangle.
\end{eqnarray}
We see that all parameters $\alpha_{jk}$ and $\beta_{jk}$ are collected, respectively,  in the amplitudes $a_j$ and phases $\varphi_j$ both defined in (\ref{tildea}).  This step terminates the state encoding up to the normalization $G$ that will be determined below after garbage removal.

Now we label and remove the garbage. To this end we  introduce the  2-qubit ancilla $B$ in the ground state $|0\rangle_B$,  the projector
\begin{eqnarray}
P_{R\varphi A } = |0\rangle_R |0\rangle_\varphi |1\rangle_{A_1}
 |1\rangle_{A_2}   \,
\, _{R}\langle 0| _\varphi\langle 0| {_{A_1}\langle 1|}  {_{A_2}\langle 1|}
\end{eqnarray}
and the controlled operator
\begin{eqnarray}\label{W4}
W^{(4)}_{R\varphi A B} = P_{R\varphi A }   \otimes \sigma^{(x)}_{B_1} \sigma^{(x)}_{B_2} +
(I_{R\varphi A}  -    P_{R\varphi A}    )\otimes  I_{B}.
\end{eqnarray}
Applying this operator to the state $|\Phi_4\rangle|0\rangle_B$ we obtain the state $|\Phi_5\rangle$, see Fig.\ref{Fig:instgen},
\begin{eqnarray}\label{Phi5}
&&
|\Phi_5\rangle =W^{(4)}_{R\varphi A B} |\Phi_4\rangle|0\rangle_B=\\\nonumber
&&
\frac{1}{2^{(n+4m)/2}}  \sum_{j=0}^{2^n-1} a_je^{2\pi i\varphi_j} |j\rangle_S |0\rangle_R|0\rangle_\varphi
 |1\rangle_{A_1}  |1\rangle_{A_2}  |1\rangle_{B_1}
|1\rangle_{B_2}  + |g_4\rangle|0\rangle_{B_1} |0\rangle_{B_2}
\end{eqnarray}
thus labeling  by $ |1\rangle_{B_1}
|1\rangle_{B_2}$ and $ |0\rangle_{B_1}
|0\rangle_{B_2}$, respectively,   the useful information and the  garbage.
Finally, we perform the controlled measurement $M_{B_2}$ over the second qubit of the ancilla $B$   via the operator
\begin{eqnarray}\label{W5}
W^{(5)}_{B} = |1\rangle_{B_1}\, {_{B_1}}\langle 1| \otimes M_{B_2} +
|0\rangle_{B_1}\,{_{B_1}\langle 0| }\otimes I_{B_2}.
\end{eqnarray}
Thus, applying $W^{(5)}_{B}$ to the state $|\Phi_5\rangle$ we obtain the state $|\Phi_6\rangle$ including the resulting state $|\Psi_{out}\rangle$, see Fig.\ref{Fig:instgen},
\begin{eqnarray}\label{Phi6}
&&
|\Phi_6\rangle = W^{(5)}_{B} |\Phi_5\rangle  = |\Psi_{out}\rangle  |0\rangle_R|0\rangle_\varphi |1\rangle_{A_1}  |1\rangle_{A_2}  |1\rangle_{B_1}  ,\\\nonumber
&&
|\Psi_{out}\rangle =G^{-1}  \sum_{j=0}^{2^n-1} a_je^{2\pi i\varphi_j} |j\rangle_S,\;\;\;
G= \sqrt{\sum_{j=0}^{2^n-1} a_j^2}.
\end{eqnarray}
We emphasize that the normalization $G$ appears in the state $|\Psi_{out}\rangle$ and coincides with that in the state
$|\Psi\rangle$ given in (\ref{Psi}).  This step concludes the state encoding algorithm.

 {\it Characteristics of algorithm.} The depth of the circuit is mainly determined by the operator $W^{(1)}_{SRA}$, whose depth is $O\Big(( 2^{n+1} n +m )m\Big)$, and by the operator $W^{(2)}_{S\varphi A_1}$, whose depth is $O\Big( 2^n( n + m )\Big)$ and can be estimated as $O(2^n n m + m^2)$.
 The space is $O(n+m)$. For $n\gg m$ the depth and space can be estimated, respectively, as $O(2^n n)$ and $O(n)$.
Both above characteristics depend on the number of qubits $n$ in the encoded superposition state (subsystem $S$)  and  the number of qubits $m$ in the auxiliary subsystems $R$ and $\varphi$.
 It is important  that the  parameter  $m$, counting the number of decimals (binary encoding) in the amplitudes and phases and fixing the dimension of the auxiliary subsystems $R$ and $\varphi$, is not related to  the number of qubits $n$ in the superposition state $|\Psi\rangle$ (\ref{Psi}) and is determined exclusively  by the required precision of state encoding.

  We give an example of a one-qubit state creation in Appendix.

{\it Conclusions.}
\label{Section:conclusions}
We propose an algorithm for creating an arbitrary quantum pure  superposition state  encoding the amplitudes and phases of this state up to certain precision.  This algorithm does not impose any special requirement on the state to be created and therefore it is the universal algorithm. It is important that our algorithm does not require additional calculations of the parameters for rotations and controlled  operations,  and therefore it does not include  supplementary classical  computations.
The required  precision determines the number of qubits in the auxiliary subsystems $R$ and $\varphi$ and the normalization factor $G$  in (\ref{Psi}).  The depth of this algorithm is
$O(2^nnm+ m^2)$ and the space is $O(n+m)$ qubits. This algorithm can be used as a subroutine in any quantum algorithm requiring creation of an initial state, for instance,    in the algorithms of matrix manipulations, developed in \cite{ZQKW_2024,ZBQKW_arXive2024}, to encode input matrices.
We remark that the parameters $n$ and $m$ are completely independent. Although the depth is  seemingly large, the proposed algorithm can be effective in the long calculation codes including set of matrix multiplications, additions and inversions.  We shall emphasize that the controlled measurement (\ref{W5}) is the  crucial  step allowing to avoid the problem of small success probability that appears if we implement  the usual measurement instead of the controlled one.

{\it Acknowledgments.}
This project is supported by the National Natural Science Foundation of China (Grants No. 12031004,
No. 12271474 and No. 61877054). The work was also partially funded by a state task of Russian Fundamental Investigations
(State Registration No. 124013000760-0).

\vspace{1cm}

\begin{figure}[ht]
  \includegraphics[width=0.6\textwidth]{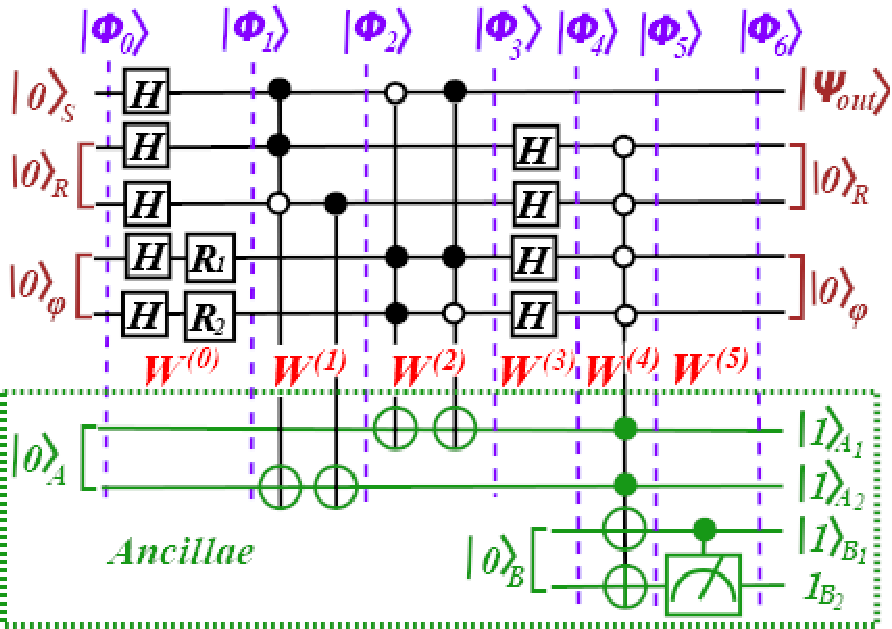}
    \caption{The circuit for creating one-qubit state (\ref{PsiRes}).}
\label{Fig:inst}
\end{figure}

{\it Appendix: Example.}
As a simple example we consider the creation of the state
\begin{eqnarray}\label{PsiRes}
|\Psi\rangle =\frac{- 2 i |0\rangle -3 |1\rangle}{\sqrt{13}}.
\end{eqnarray}
The circuit for this example is presented in Fig.\ref{Fig:inst}.
In this case $n=1$, $m=2$, $\alpha_{00}=0$, $\alpha_{01}=\alpha_{10}=\alpha_{11}=1$, $\beta_{01}=\beta_{02}=\beta_{11}=1$, $\beta_{12}=0$,
$R_{\varphi_1}={\mbox{diag}}(1,-1)$, $R_{\varphi_2}={\mbox{diag}}(1,i)$.
 Formula (\ref{Phi1}) for $|\Phi_1\rangle$ now reads
\begin{eqnarray}
|\Phi_1\rangle &=& \frac{1}{2^{5/2}} (|0\rangle_S + |1\rangle_S)(|00\rangle_{R} + |01\rangle_{R}+|10\rangle_{R} + |11\rangle_{R}) |\Psi_\varphi\rangle, \\\nonumber
&&
 |\Psi_\varphi\rangle = (|0\rangle_{\varphi_1} - |1\rangle_{\varphi_1})(|0\rangle_{\varphi_2} +i |1\rangle_{\varphi_2}).
\end{eqnarray}
{\it Amplitude encoding.}
For the projectors (\ref{P2}) we have
 $P^{(0)}_{S} =  |1\rangle_S\, {_S\langle 1|}$,
$P^{(1)}_{S} = |0\rangle_S\, {_S\langle 0|} +  |1\rangle_S\, {_S\langle 1|} \equiv I_S$.
Therefore, (\ref{tW1}) yields
\begin{eqnarray}\label{ExW1}
 \tilde W^{(0)}_{SA_1} =  |1\rangle_S\, {_S\langle 1|}\otimes \sigma^{(x)}+  |0\rangle_S\, {_S\langle 0|}\otimes I_{A_1} ,\;\; \tilde W^{(1)}_{SA_1} = I_{S} \otimes \sigma^{(x)}_{A_1}.
\end{eqnarray}
Next, projectors (\ref{P1}) read
 $P^{(0)}_{RA_1} =|1\rangle_{R_1}|0\rangle_{R_2}\, {_{R_1}\langle 1|}  {_{R_2}\langle 0|} \otimes |1\rangle_{A_1}\,{_{A_1}\langle 1|}$,
$P^{(1)}_{RA_1} =|1\rangle_{R_2}\, {_{R_2}\langle 1|} \otimes |1\rangle_{A_1}\,{_{A_1}\langle 1|}$,
Then,  formula (\ref{W1}) yeilds
\begin{eqnarray}\label{ExV1}
&&V^{(0)}_{RA} = P^{(0)}_{RA_1}\otimes \sigma^{(x)}_{A_2}  + (I_{RA_1}- P^{(0)}_{RA_1})\otimes I_{A_2},\\\nonumber
&&V^{(1)}_{RA} = P^{(1)}_{RA_1}\otimes \sigma^{(x)}_{A_2}  + (I_{RA_1}- P^{(1)}_{RA_1})\otimes I_{A_2}.
\end{eqnarray}
The operators $\tilde W^{(k)}_{SA_1}$ (\ref{ExW1}) and $V^{(k)}_{RA}$   (\ref{ExV1}) can be combined in formula (\ref{W12}) as follows, see operator $W^{(1)}$ in Fig.\ref{Fig:inst}:
\begin{eqnarray}\nonumber
\tilde W^{(0)}_{SA_1} V^{(0)}_{RA} \tilde W^{(0)}_{SA_1} &=& P_{SR}\otimes \sigma^{(x)}_{A_2}  + (I_{SR}- P_{SR})\otimes I_{A_2},\\\nonumber
&& P_{SR}= |1\rangle_S\,_S\langle 1| \otimes|1\rangle_{R_1}|0\rangle_{R_2}\, {_{R_1}\langle 1|}  {_{R_2}\langle 0|},\\\nonumber
\tilde W^{(1)}_{SA_1} V^{(1)}_{RA}\tilde W^{(1)}_{SA_1}  &=&|1\rangle_{R_2} \, {_{R_2}\langle 1|} \otimes \sigma^{(x)}_{A_2}  +|0\rangle_{R_2} \, {_{R_2}\langle 0|} \otimes I_{A_2}.
\end{eqnarray}
Then (\ref{Phi2}) reads
\begin{eqnarray}\nonumber
&&
|\Phi_2\rangle =\frac{1}{2^{5/2}}\Big( |0\rangle_S \big(|10\rangle_R + |11\rangle_R\big)   +
 |1\rangle_S \big(|01\rangle_R +|10\rangle_R + |11\rangle_R\big)
 \Big)|\Psi_\varphi\rangle |10\rangle_{A}\\\nonumber
 &&+|g_2\rangle
\end{eqnarray}
{\it Phase encoding.} Now we start creating the proper phases of the probability amplitudes.
From (\ref{P22}) we have
$P^{(0)}_{S\varphi} = |0\rangle_S \, {_S\langle 0|} \otimes |11\rangle_\varphi \,{_\varphi\langle 11|}$,
$P^{(1)}_{S\varphi} =  |1\rangle_S \, {_S\langle 1|} \otimes |01\rangle_\varphi \,{_\varphi\langle 01|}$.
Then eq.(\ref{Phi3}) in view of operators $\tilde W^{(j)}_{S\varphi A_1}$  (\ref{W2SA}) and $W^{(2)}_{S\varphi A_1}$  (\ref{W2SA2}) yields
\begin{eqnarray}
&&|\Phi_3\rangle = \frac{1}{2^{5/2}}
\Big( |0\rangle_S \big(|10\rangle_R + |11\rangle_R\big) (-i) |11\rangle_{\varphi} \\\nonumber
&& +
 |1\rangle_S \big(|01\rangle_R +|10\rangle_R + |11\rangle_R\big)(-1)  |01\rangle_{\varphi }
 \Big) |11\rangle_{A}+|g_3\rangle
\end{eqnarray}
After applying $W^{(3)}_{R\varphi}$ (\ref{W3}), we obtain $|\Phi_4\rangle$ (\ref{Phi4}):
\begin{eqnarray}
|\Phi_4\rangle &=& \frac{1}{2^{9/2}}
\Big( -2i |0\rangle_S  -3
 |1\rangle_S
 \Big)|00\rangle_{R}  |00\rangle_{\varphi }  |11\rangle_{A}+|g_4\rangle.
\end{eqnarray}
 Including the  two-qubit ancilla $B$, labeling the  garbage via the controlled operator  $W^{(4)}_{R\varphi A B}$ (\ref{W4}) and removing the garbage applying the controlled  measurement $W^{(5)}_{B}$ (\ref{W5})  we obtain  the final state $|\Phi_6\rangle$ (\ref{Phi6})  where  $|\Psi_{out}\rangle $ equals the required state $|\Psi\rangle$ given  in (\ref{PsiRes}).

\end{document}